# Bisecting for selecting: using a Laplacian eigenmaps clustering approach to create the new European football Super League

A. J. Bond[1] & C. B. Beggs[1]

[1]Institute for Sport, Physical Activity and Leisure, Carngergie School of Sport, Leeds Beckett University, Leeds, West Yorkshire, United Kingdom.

**Alexander John Bond (Corresponding Author)**

215 Cavendish Hall, Headingley Campus, Leeds Beckett University, LS6 3QS.

A.J.Bond@leedsbeckett.ac.uk

Clive Beggs,

Faifax Hall, Headingley Campus, Leeds Beckett University, LS6 3QS.

C.Beggs@leedsbeckett.ac.uk

# Bisecting for selecting: using a Laplacian eigenmaps clustering approach to create the new European football Super League

**Abstract**

We use European football performance data to select teams to form the proposed European football Super League, using only unsupervised techniques. We first used random forest regression to select important variables predicting goal difference, which we used to calculate the Euclidian distances between teams. Creating a Laplacian eigenmap, we bisected the Fielder vector to identify the five major European football leagues' natural clusters. Our results showed how an unsupervised approach could successfully identify four clusters based on five basic performance metrics: shots, shots on target, shots conceded, possession, and pass success. The top two clusters identify those teams who dominate their respective leagues and are the best candidates to create the most competitive elite super league.

1. **Introduction**

Operational research (OR) has a long history of using sport to explore operational insights and methodologies (see Wright, 2009 for a review). Recently association football (herein football, also known as soccer), has become a popular data source to understand a range of operational issues like scheduling (Durán et al., 2017; Yi et al., 2020); rating and valuing (Baker & McHale, 2018; Kharrat et al., 2020; Müller et al., 2017); managerial succession (Beggs & Bond, 2020; Flores et al., 2012); events planning (Ghoniem et al., 2017); performance prediction (Beggs et al., 2019; Butler et al., 2021), to name a few. While data availability makes football attractive to operational researchers, football needs operational research to drive decision-making. For example, currently, there are substantial plans to disrupt European football operations to introduce a new elite European Super League similar to the EuroLeague established in basketball (West, 2018). Indeed, this is not a new proposal, it has rumbled under the surface since the nineties, but with a reported $6 billion funding package from investment firm JP Morgan Chase, it looks more likely (Conn, 2021; McInnes, 2020).

With annual revenues of approximately $28 billion per annum (Deloitte, 2020), European football (or soccer) presents a world of high finance. In addition to high revenue-generating capacity, it is considered 'recession proof' (Dimitropoulos et al., 2016). While the entertainment industry deals with significant losses throughout the COVID-19 pandemic, European football performed reasonably well with an estimated $26 billion revenue, losing roughly twelve per cent (Deloitte, 2021). European football's sustained revenue-generating capacity predominantly stems from the global demand for the five major leagues; English Premier League, Spanish La Liga, German Bundesliga, Italian Serie A and French Ligue 1 – contributing nearly 60% of European football's total revenue (Deloitte, 2020). The global demand for these football properties is demonstrated through the large broadcasting and media rights sales, with the English Premier League dominating the market attracting €3.5 billion, compared to La Liga's €1.8 billion, Bundesliga's and Serie A's €1.5 billion, and Ligue 1's €900 million (*ibid.*). Similarly, Europe's governing body Union of European Football Associations (UEFA), generated €2 billion in 2017 from their flagship Champions League event (Global Data, 2021). Unsurprisingly, European football's revenue-generating capabilities have attracted considerable foreign investments (Rohde & Breuer, 2016; Wilson et al., 2013). Consequently, it is unsurprising that the dominant teams across Europe are looking to capitalise on their global demand, nor is it surprising that investment firms like JP Morgan are reportedly ready to invest.

If successful, this would drastically disrupt European football operations, redefining structures, revenues and operating systems. Of course, such a proposal is not backed by UEFA (Conn, 2019), or the European domestic leagues (Telegraph Sport, 2018), who would be in direct competition for consumers attention. While little is known about the actual plans being considered, the proposal seems to be eleven founder clubs; Real Madrid, Barcelona, Manchester United, Juventus, Chelsea, Arsenal, Paris Saint-Germain, Manchester City, Liverpool, AC Milan and Bayern Munich; and five initial guests; Atlético Madrid, Borussia Dortmund, Marseille, Inter Milan and AS Roma (Der Spiegel, 2018). Granted, these teams represent the elite clubs with superior financial resources. However, entry into this new proposed market seems arbitrary, based on being selected by this new league's proposers. Fundamentally, sports firms (and leagues) sell competition to consumers (Andreff & Scelles, 2015; Mackreth & Bond, 2020; Neale, 1964), so the success of a league is often associated to the competitive balance within it (Bond & Addesa, 2019, 2020; Caruso et al., 2019).

Therefore, selecting the right clubs to form a new league is imperative to a successful league and ensure a return on any investments. While there are numerous ways to measure competitiveness, these are often based on win percentages or points allocations (see Lee et al., 2019a, 2019b; Owen & King, 2015). More importantly, they mainly provide an overall measure to make comparisons over time or compare other sports leagues and provide little information about the 'who' is dominant in the market. Therefore, we propose a different, unsupervised data-driven approach to identify 'who' dominates the European leagues. Indeed, there are several attempts to classify or rank European football teams, such as; the Euro Club Index (EuroClubIndex.com, 2021), the ClubElo index (clubelo.com, 2021), and the UEFA club coefficient rankings (UEFA, 2021). However, these tend to rely on either established rating systems such as the Elo system (Elo, 1978), or the awarding of arbitrary points for progression in cup competitions. They reveal nothing about the natural groupings (clusters) that might exist in European football. Our approach is to blend machine learning and graph theory approaches to identify natural clusters of teams in Europe's five major leagues, using simple performance data. Not only does this provide an innovative approach to sport operations, allowing a more objective approach to selecting teams into the new league, but it also offers a methodology to allow operational researchers to identify natural clusters of firms within markets.

## 2. Methods

### 2.1. Data acquisition

Using publicly available football performance data from footystats.com (FootyStats.com, 2021) and WhoScored.com (WhoScored.com, 2021), we used simple performance data for all the teams in the Bundesliga, La Liga, Ligue 1, English Premier League, and Serie A over seven seasons between 2013/14 – 2019/2020. This produced a study data set comprising 686 observations from a total of 150 football teams. The variables collected related to an individual team's performance over the entire season. The teams are listed in Appendix 1, and the variables included in the study are listed in Table 1. The data from all seven seasons were aggregated for each team into a single dataset (n = 150) to avoid pseudoreplication. This aggregated dataset was used to perform the data analysis.

Table 1. Variable description and descriptive statistics

| Variable | Description | Mean | SD | Median | Min | Max |
|---|---|---|---|---|---|---|
| Yellow_cards | Number of yellow cards received | 75.7 | 17 | 70.79 | 43.7 | 116 |
| Red_cards | Number of red cards received | 4.06 | 1.76 | 4 | 0.5 | 9 |
| Possession | Possession percentage | 48.8 | 4.2 | 47.81 | 39.1 | 64.14 |
| Pass_Success | Successful pass percentage | 77.2 | 4.48 | 76.9 | 62.1 | 89.09 |
| Aerials_Won | Number aerial duals won | 18.2 | 3.79 | 17.66 | 9.8 | 30.65 |
| Shots_Conceeded | Number shots conceded per game | 13.1 | 1.97 | 12.87 | 8.04 | 18.55 |
| Tackles | Number of tackles made per game | 18.2 | 1.64 | 18.33 | 13.3 | 23 |
| Interceptions | Number of interceptions made per game | 14.3 | 2.19 | 14.14 | 9.5 | 22.3 |
| Fouls | Number of fouls conceded per game | 13.3 | 1.7 | 13.51 | 9.34 | 16.9 |
| Offsides | Number of offsides per game | 2.1 | 0.38 | 2.07 | 1.25 | 3.4 |
| Shots | Number of shots per game | 12.2 | 1.74 | 11.83 | 8.8 | 17.61 |
| Shots_OT | Number of shots on target per game | 4.13 | 0.84 | 3.95 | 2.6 | 7.03 |
| Dribbles | Number of dribbles made per game | 9.12 | 1.73 | 9.09 | 4.75 | 14.1 |
| Fouled | Number of time fouled by opposing team | 12.5 | 1.67 | 12.74 | 7.93 | 17.1 |
| GF | Goals scored | 46.3 | 14 | 42.58 | 22 | 101.9 |
| GA | Goals conceded | 54.3 | 11.7 | 54.89 | 24.6 | 85 |
| GD | Goal difference (GF-GA) | -8.06 | 23.4 | -12.98 | -51 | 70 |
| Points | Total points gained | 45.6 | 15.1 | 42.36 | 15 | 91.29 |

### *2.2. Data analysis strategy*

The study aimed to develop a methodology for identifying natural groupings between teams in the various European soccer leagues, using season match data alone (excluding goals scored or conceded). We performed an exploratory analysis using basic univariate analysis on the variables used in this study before conducting a random forest regression analysis to identify the measured variables that best predicted the goal difference for the respective soccer teams. Goal difference was used because it is a better measure of team performance and less susceptible to bias than points total, which is influenced by the number of teams in the respective leagues (Heuer & Rubner, 2009).

Having identified the variables that best predicted end-of-season goal difference, we then computed the Euclidean distances between the respective teams in the vector space and used them to produce Laplacian eigenmaps of the data (Belkin & Niyogi, 2003). Laplacian eigenmaps are constructed from the eigenvectors of a graph Laplacian matrix. They are essentially an embedding algorithm that seeks to project pair-wise proximity information onto a low dimensional space to preserve local structures in the data. Unlike linear dimension reduction techniques such as principal component analysis (PCA), Laplacian eigenmaps have the great advantage they can handle non-linear relationships in the data (Belkin & Niyogi, 2003; Nascimento & de Carvalho, 2011). Therefore by producing Laplacian eigenmaps, we succinctly visualise the relationships between the respective soccer teams and identify sub-groups within the data using spectral cluster analysis techniques. To benchmark our findings, we classified the respective teams according to their points, using 25% and 75% percentiles to reflect top and bottom performing teams, otherwise classed as middle teams. The 25% and 75% percentiles turned out to be >56 points classified top teams, <36 points classified bottom teams, with all others classified as middle. All data and statistical analysis were performed using in-house algorithms written in R (R Core Team, 2020).

### *2.3. Statistical analysis*

An initial univariate analysis of the aggregated data was undertaken using a one-way ANOVA, with post-hoc Bonferroni adjusted pair-wise t-tests. This allowed a better understanding of the data and variables used in this study.

### 2.3.1. Exploratory random forest analysis

An exploratory random forest regression was performed to assess the observed variables' relative importance as predictors of goal difference. Random forest analysis is an ensemble classification technique popular in machine learning that generalises classification trees (Boinee et al., 2008; Breiman, 2001). It is a robust technique resistant to over-fitting and does not require strict distributional assumptions (Breiman, 2001; Izenman, 2013). Crucially, it has the great advantage that it can assess variable importance, thus enabling the user to discard redundant variables that do not assist in the prediction process.

Random forest models produce many regression trees that use recursive partitioning of data to group observations into predefined classes through binary splitting the predictor variables (Hansen et al., 2015). Bias and over-fitting are minimised by employing a combination of bootstrap bagging and utilising a random subset of predictor variables (generally the square root of the total number of predictors in the model) at each split. Each regression tree in the random forest is built using a bootstrapping algorithm, which randomly 'bags' a sample from approximately two-thirds of the data for training purposes, leaving the remaining one-third of the cases or out-of-bag (OOB) cases to assess the performance of the regression tree (Boinee et al., 2008; Cutler et al., 2007). For each tree, the prediction error – mean squared error (MSE) in the case of a regression tree – is computed. These are then pooled to give an overall measure of classification accuracy, thus ensuring that the assessment is unbiased (Pecl et al., 2011).

We used the 'randomForest' package (Liaw & Wiener, 2002) in R (R Core Team, 2020) to perform a random forest analysis involving the creation of 500 random trees. Initial analysis was undertaken using all thirteen predictor variables to identify those variables that significantly influenced the outcome variable, Goal_Difference. The 13 predictor variables used to predict goal difference were; shots on target; possession; shots; shots conceded; pass_success; dribbles; aerials won; offsides; tackles; yellow cards; red cards; fouls; fouled; interceptions, described in Table 1. The relative importance of the variables was assessed using the Gini variable importance measure (VIM), which we corrected for bias using the heuristic strategy proposed by Sandri and Zuccolotto (2008, 2010) and implemented by Carpita et al. (2014). For every node split in a tree, the Gini impurity criterion (which assesses the data's heterogeneity) for the two descendent nodes is less than that of the parent node (Friedman, 2001). Therefore, adding up the Gini decreases for each variable over all trees in the forest, it is possible to achieve a measure of variable importance. In our analysis, variables that exceeded the inflexion point's value on the Gini VIM curve were deemed to be influential and thus retained when the random forest model was refined. Having identified the key variables that best predicted goal difference, we then repeated the random forest analysis using the refined model to understand the prediction accuracy that could be achieved. Prediction of the respective teams' goal differences was then performed using the refined model and an ensemble prediction algorithm that aggregated 500 predictions. Because random forests use a self-validating MSE rate, there is strictly no need for cross-validation or a separate validation test to obtain an unbiased estimate of model error (Pecl et al., 2011). However, to demonstrate the refined random forest model's validity, we performed k-fold cross-validation using ten randomly sampled 'folds' of approximately equal size.

*2.3.2. Laplacian eigenmaps*

Spectral cluster analysis was performed using a Laplacian eigenmaps method to visualise relationships between the respective teams and identify natural sub-groups within the data, as described by Belkin & Niyogi (2003). This approach involves computing the pair-wise Euclidean distances between the respective teams using the key variables identified by the random forest analysis. These were transformed into a [150 x 150] similarity matrix, $Q$, using a Gaussian radial basis function (rbf) kernel (Schölkopf et al., 2004), with $\sigma = 1$, as follows:

$$Q = exp\left(-\frac{E^2}{2 \times \sigma^2}\right) \quad (1)$$

where; $E$ is the matrix of pair-wise Euclidean distances. The non-linear Gaussian function filtered the Euclidean distance matrix so that edges between close neighbours were given more weight compared with those between teams more distantly separated. From this, the modified similarity matrix, $W$, was constructed by subtracting the [150 x 150] identify matrix, $I$, from the similarity matrix, $Q$:

$$W = Q - I \quad (2)$$

This was then be used to construct the degree matrix, $D$, as follows:

$$s = W.n \quad (3)$$

where, $n$ is a [150 x 1] vector of ones and $D$ is:

$$D_{ij} = \begin{cases} s_i & if\ i = j \\ 0 & if\ i \neq j \end{cases} \quad (4)$$

Having computed the degree matrix, $D$, the Laplacian, $L$, and normalised Laplacian, $L_{norm}$, matrices (both symmetric, positive semi-definite matrices) were then constructed (Chung & Graham, 1997; Qiu & Hancock, 2004; von Luxburg, 2007), as follows:

$$L = D - W \quad (5)$$

$$L_{norm} = D^{-0.5}.L.D^{-0.5} \quad (6)$$

After this, eigendecomposition of the normalised Laplacian matrix, $L_{norm}$, was performed in order to compute

the diagonal matrix of eigenvalues, Λ, and the matrix of eigenvectors, *V*, as follows:

$$L_{norm} = V.\Lambda.V^T$$

(7)

However, unlike PCA, where the eigenvectors corresponding to the largest eigenvalues are used to construct the principal components, Laplacian eigenmaps construct a configuration from the eigenvectors corresponding to the two or three smallest positive eigenvalues. Because the smallest eigenvalue equals zero, the eigenvector corresponding to this eigenvalue is often ignored, and instead, the eigenvectors associated with the successive two or three smallest positive eigenvalues are used to construct the map (Qiu & Hancock, 2004). In our case, we used the last three positive eigenvectors, fourth, third and second (Fielder) smallest eigenvectors, to construct 3D Laplacian eigenmaps of the European football teams. We used third and Fielder vectors to construct 2D Laplacian eigenmaps.

### *2.3.3. Natural Clustering Approach*

Laplacian eigenmaps are considered to be a spectral clustering technique. As such, it exhibits a critical property first discovered by Fiedler (1975), namely that the eigenvector associated with the second smallest eigenvalue (i.e. the smallest positive eigenvalue) can be used to partition a graph. The Fiedler vector, as it has is known, is widely used in spectral graph partitioning (Higham et al., 2007; Naumov & Moon, 2016; Stone & Griffing, 2009) as an unsupervised technique for bisecting graphs, enabling sub-groups (clusters) within the data to be readily identified. Multiple sub-groups can be identified by repeated bisection of the Laplacian eigenmaps using the Feidler vector (Naumov & Moon, 2016).

To identify how many bisections were appropriate to establish the natural clusters in the data, we ran a cluster validation using the 'clValid' package in R (Brock et al., 2008). To do so, we used the self-organising maps algorithm (Kohonen, 1991, 2012) since it is an unsupervised learning technique partitioning data using artificial neural networks. To determine the suitability of 2 – 6 partitions of the fielder vector, internal consistency was measured by the Dunn Index (Dunn, 1974) and Silhouette Width (Rousseeuw, 1987), both of which should be maximised (see Handl et al., 2005, for a review). The Silhouette Widths were also used to inspect final cluster classifications, following the Fielder vector's bisection. To visualise natural clustering, we created an undirected graph network using the inverse of the Euclidean distances between the respective teams.

### 3. Results
### *3.1. Descriptive Statistics*

The descriptive analysis results using the aggregated data split by benchmark percentiles (top, middle, bottom) and the one-way ANOVA are presented in Table 1. Unsurprisingly, the top teams had significantly greater possession and pass success; conceded fewer shots; made more dribbles and shots than weaker teams (all

p<0.001); had greater possession and pass success (both p<0.001); and made more dribbles and shots (both p<0.001), compared with the weaker teams. In addition, they made significantly fewer fouls (p = 0.037) but did not significantly receive less yellow (p = 0.214) and red (p = 0.406) cards.

Table 1. Descriptive statistical results for aggregated data (all seasons) together with the one-way ANOVA results.

|  | Bottom (N=37) Mean (SD) | Middle (N=81) Mean (SD) | Top (N=32) Mean (SD) | Total (N=150) Mean (SD) | ANOVA Sig. | Pair-wise Significant Differences (p = < .05) |
|---|---|---|---|---|---|---|
| Yellow_cards | 78.379 (16.887) | 76.219 (17.647) | 71.335 (15.232) | 75.710 (17.041) | 0.214 | Not Sig. |
| Red_cards | 4.157 (1.848) | 4.158 (1.812) | 3.685 (1.510) | 4.057 (1.761) | 0.406 | Not Sig. |
| Possession | 46.090 (2.575) | 47.693 (2.525) | 54.557 (3.815) | 48.762 (4.203) | < 0.001 | 1,2,3 |
| Pass_Success | 75.221 (4.033) | 76.082 (3.356) | 82.464 (3.470) | 77.231 (4.482) | < 0.001 | 1,2,3 |
| Aerials_Won | 18.209 (4.623) | 18.901 (3.386) | 16.223 (3.069) | 18.159 (3.793) | 0.003 | 3 |
| Shots_Conceeded | 14.958 (1.653) | 13.146 (1.360) | 10.850 (1.146) | 13.103 (1.967) | < 0.001 | 1,2,3 |
| Tackles | 18.481 (2.373) | 18.114 (1.279) | 18.233 (1.430) | 18.230 (1.639) | 0.532 | Not Sig. |
| Interceptions | 14.177 (2.431) | 14.542 (2.124) | 13.606 (2.000) | 14.252 (2.195) | 0.12 | Not Sig. |
| Fouls | 13.591 (1.800) | 13.482 (1.655) | 12.656 (1.569) | 13.333 (1.701) | 0.037 | 2,3 |
| Offsides | 2.019 (0.441) | 2.048 (0.306) | 2.330 (0.392) | 2.101 (0.379) | < 0.001 | 2,3 |
| Shots | 11.380 (1.231) | 11.651 (0.970) | 14.533 (1.791) | 12.199 (1.743) | < 0.001 | 2,3 |
| Shots_OT | 3.617 (0.480) | 3.888 (0.447) | 5.325 (0.809) | 4.128 (0.838) | < 0.001 | 1,2,3 |
| Dribbles | 8.666 (2.052) | 8.745 (1.292) | 10.586 (1.508) | 9.118 (1.725) | < 0.001 | 2,3 |
| Fouled | 12.863 (1.988) | 12.348 (1.616) | 12.464 (1.353) | 12.500 (1.668) | 0.298 | Not Sig. |
| GF | 35.137 (5.426) | 43.007 (5.643) | 67.335 (13.671) | 46.256 (13.963) | < 0.001 | 1,2,3 |
| GA | 66.290 (10.102) | 54.662 (6.296) | 39.601 (6.200) | 54.317 (11.667) | < 0.001 | 1,2,3 |
| GD | -31.153 (9.924) | -11.655 (8.944) | 27.734 (17.872) | -8.062 (23.405) | < 0.001 | 1,2,3 |
| Points | 29.614 (5.280) | 43.591 (4.988) | 69.025 (10.534) | 45.569 (15.056) | < 0.001 | 1,2,3 |

Legend:

1. Significant after Bonferroni adjustment between Bottom and Middle.

2. Significant after Bonferroni adjustment between Bottom and Top.

3. Significant after Bonferroni adjustment between Middle and Top.

### *3.2. Random Forest Analysis Results*

The exploratory random forest analysis incorporating all the predictor variables produced a regression model with an MSE of 115.62 and an $r^2$ value of 0.7875 (or 78.75% explained variance), which was used to assess variable importance (see Figure 1). From Figure 1, it can be seen that the Gini VIM values for the five variables: Shots_OT (on target); Possession; Shots_conceded; Shots; and Pass_Success, were far more than the values for the other variables, which were subsequently discarded from the refined random forest regression model. As such, this indicates that these five variables were the best predictors of end-of-season goal difference.

Figure 1. Random forest regression Gini corrected VIM

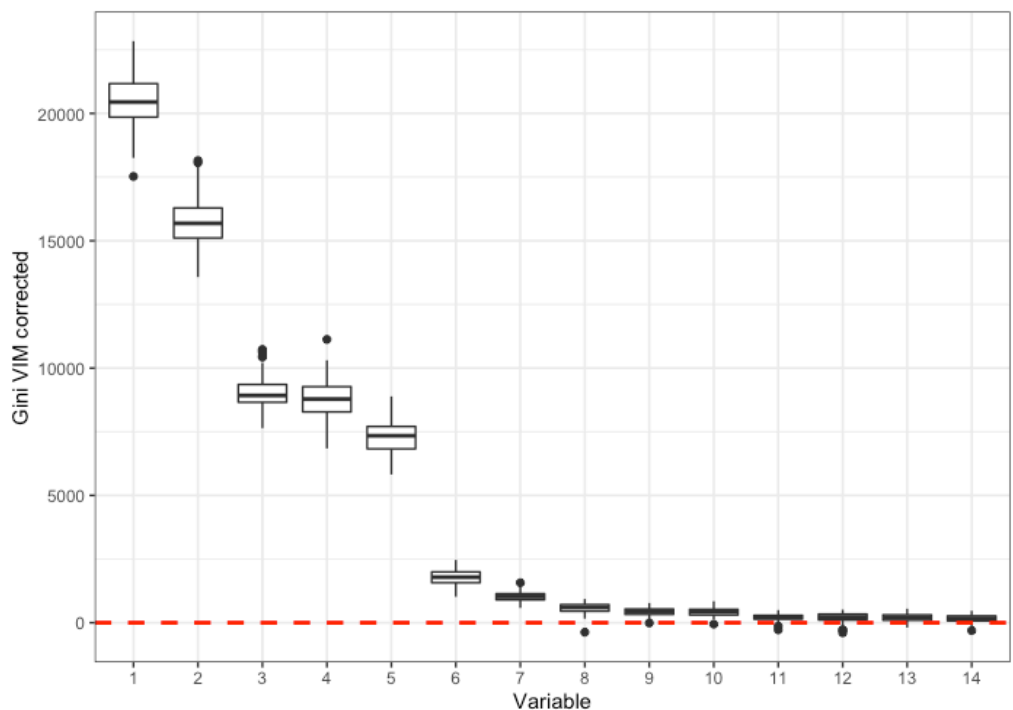

Legend: (1) Shots_OT; (2) Possesion; (3) Shots; (4) Shots_Conceded; (5) Pass_Success; (6) Dribbles; (7) Arials_Won; (8) Offsides; (9) Tackles; (10) Yellow_Cards; (11) Red_Cards; (12) Fouled; (13) Interceptions; (14) Fouls

The refined random forest analysis utilising only these important variables produced a regression model with an MSE of 113.84 and an $r^2$ value of 0.7908 (79.08% variance explained). This was confirmed by the 10-fold cross-validation process, which found the cross-validation MSE to be 113.44, with a cross-validation $r^2$ value of 0.7908 (79.08% variance explained. The relationship between predicted and actual goal difference for the respective clubs is shown in Figure 2. From this, it can be seen that the refined random forest model predicted the end-of-season goal difference with a high degree of accuracy.

Figure 2. Scatter plot of predicted goal difference versus actual goal difference for the refined random forest regression model

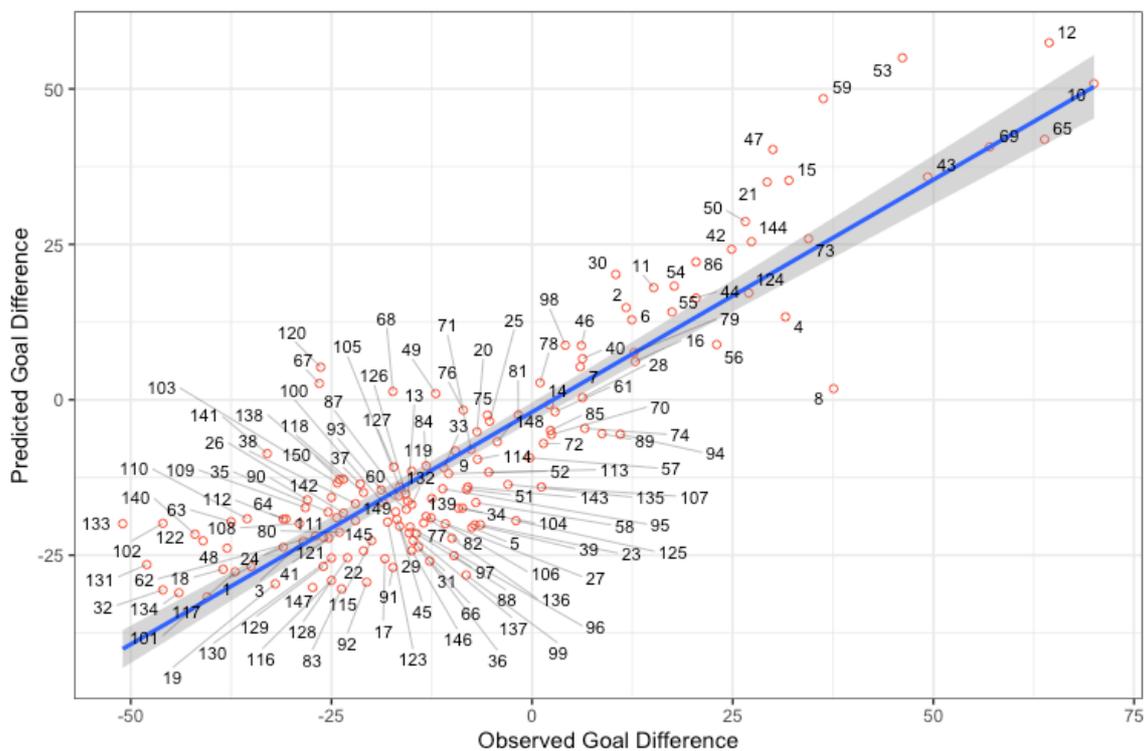

### 3.3. Laplacian Eigenmap Results

The 3D Laplacian eigenmaps of the teams are presented in Figure 3, which shows a scatter plot of the three smallest positive eigenvectors. The 3D plots demonstrate a spiral-like curve between the three dimensions, demonstrating a hierarchal structure. Figure 4 shows the 2D Laplacian eigenmap with the Fielder vector plotted against the third smallest eigenvector. Here it shows a characteristic U-shaped curve, with the teams distributed along its length. Figure 4 the teams are classified according to the 25% and 75% percentile points benchmark groupings. From this, it is relatively clear that most top clubs plot to the Feilder vector's right (>0.1), with a relatively clear distinction from the rest. Similarly, the bottom clubs tend to plot to the left of the Fielder vector (<0). However, middle clubs have a less clear space along the curve. Interestingly, La Palmas (Team No. 120; La Liga), who were benchmarked bottom, and Nice (Team No. 61; Ligue 1) plot closer to the top benchmarked teams >0.1 on the Fielder vector.

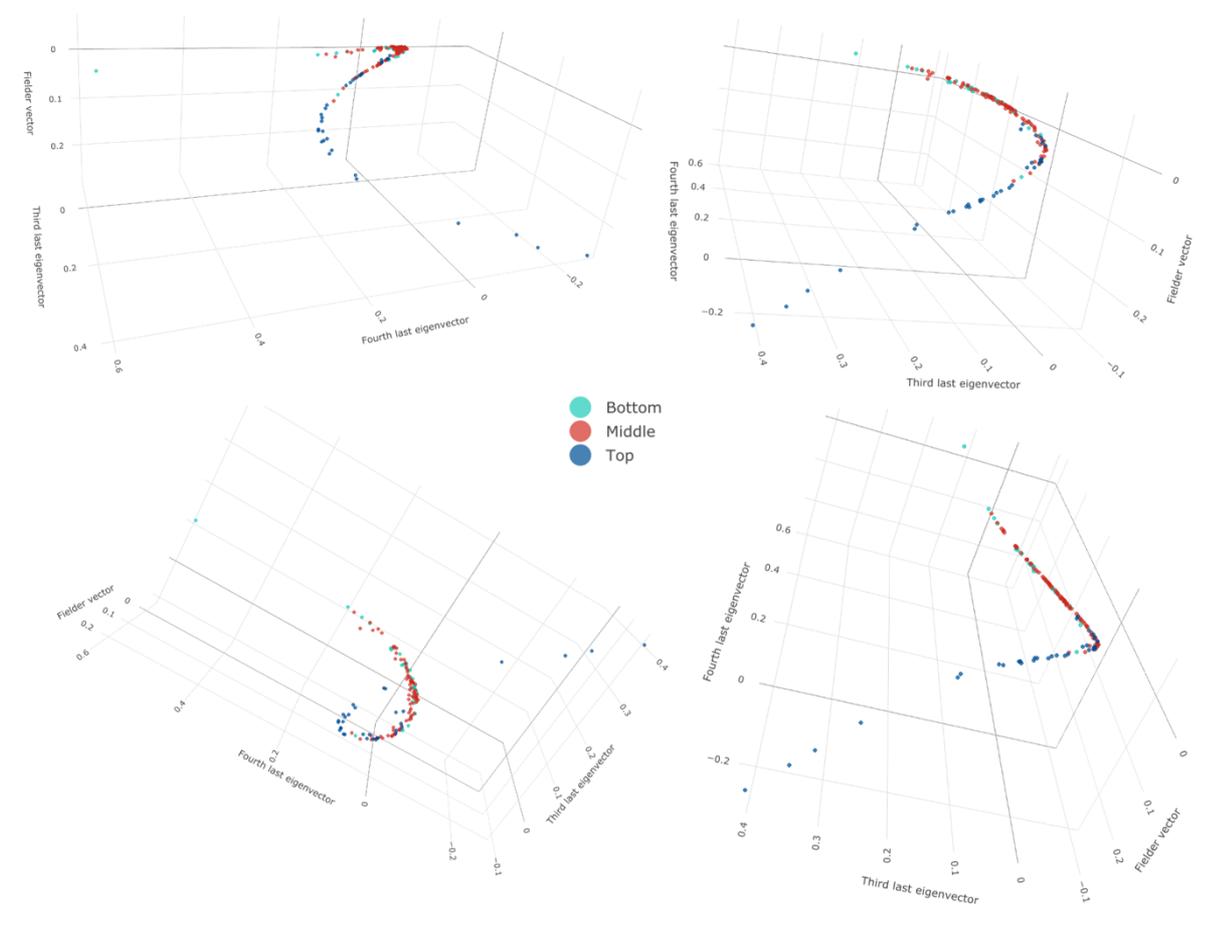

Figure 3. 3D scatter plot of Laplacian eigenmap using the three smallest positive eigenvectors (NB. The plot is viewed from four different angles).

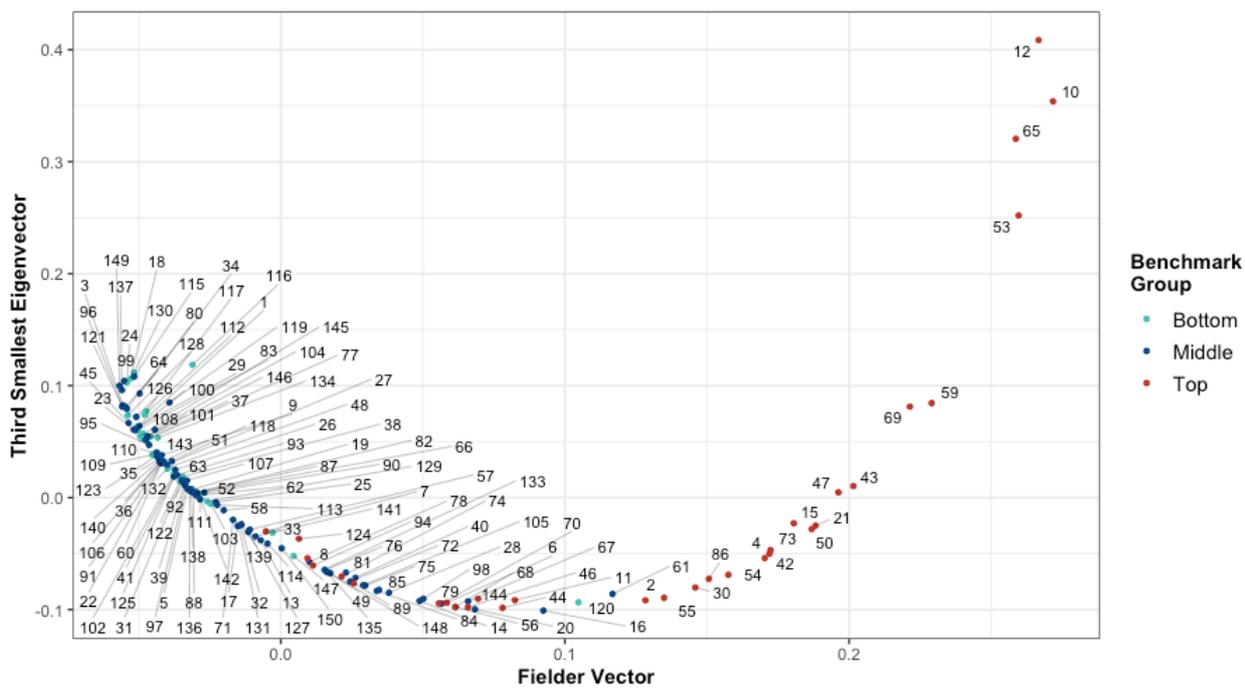

Figure 4. 2D scatter plot of Laplacian eigenmap using the Fielder vector and the third smallest eigenvector

When the benchmarked classifications are mapped onto a network graph of the inverse Euclidean distances (Figure 5), it can be seen that although top teams cluster to the bottom right, there is considerable overlap across all top, middle and bottom teams. Indeed, the average silhouette width values for the benchmark classifications were only 0.04, indicating that classification based on the national leagues' points does not accurately reflect the natural groupings between the various soccer clubs in Europe.

Figure 5. Network graph of the benchmark clusters

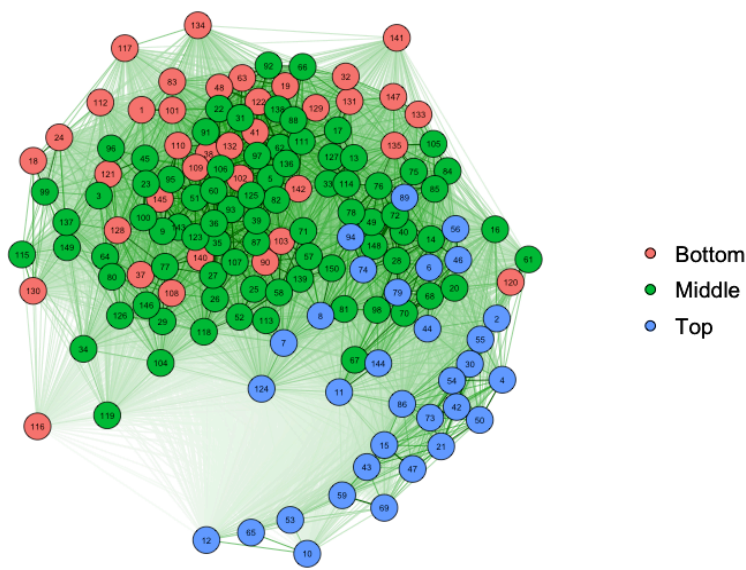

### 3.4. Natural Clustering Results

The Dunn Index and Silouhette Width results for the self-organising maps cluster validation are presented in Figure 6. It is clear that the 4 cluster solution maximises both internal validation measures, therefore, requiring three bisections of the Fielder vector.

Figure 6. Dunn Index (left) and Silhouette Width (right) cluster validation for 2 to 6 clusters using self-organising maps algorithm.

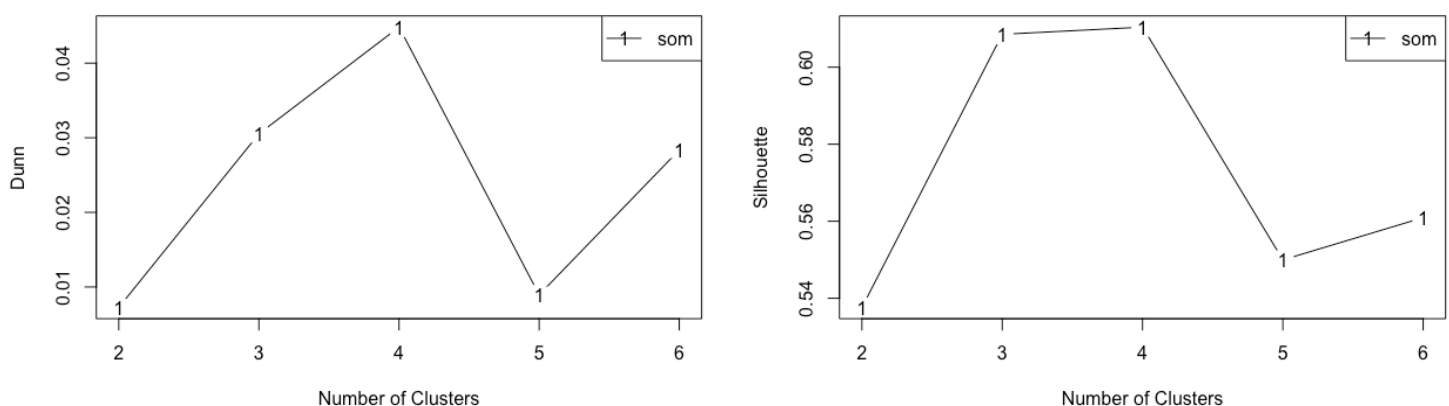

The three bisections of the Fielder vector are presented in Figure 7, creating 4 clusters SC1-SC4. Here the clusters demonstrate a group of four very strong dominating teams (SC1), fifteen strong teams (SC2), thirty-seven medium-strength teams (SC3) and ninety-four weaker teams (SC4). Overall, the natural clusters identified by the Fielder vector algorithm are well defined, with an average Sillouhette Width = 0.61, and no cluster below 0.50 (Figure 8 and Table 2), and a Dunn Index = 0.0098. The lowest internally valid cluster is SC2 with a Silhouette Index = 0.50, suggesting this group is more heterogeneous than homogeneous. The natural clustering network graph is visualised in Figure 9, which shows how cohesive the clusters are based on the inverse Euclidian distance.

Figure 7. Natural clusters from bisections of the Fielder vector

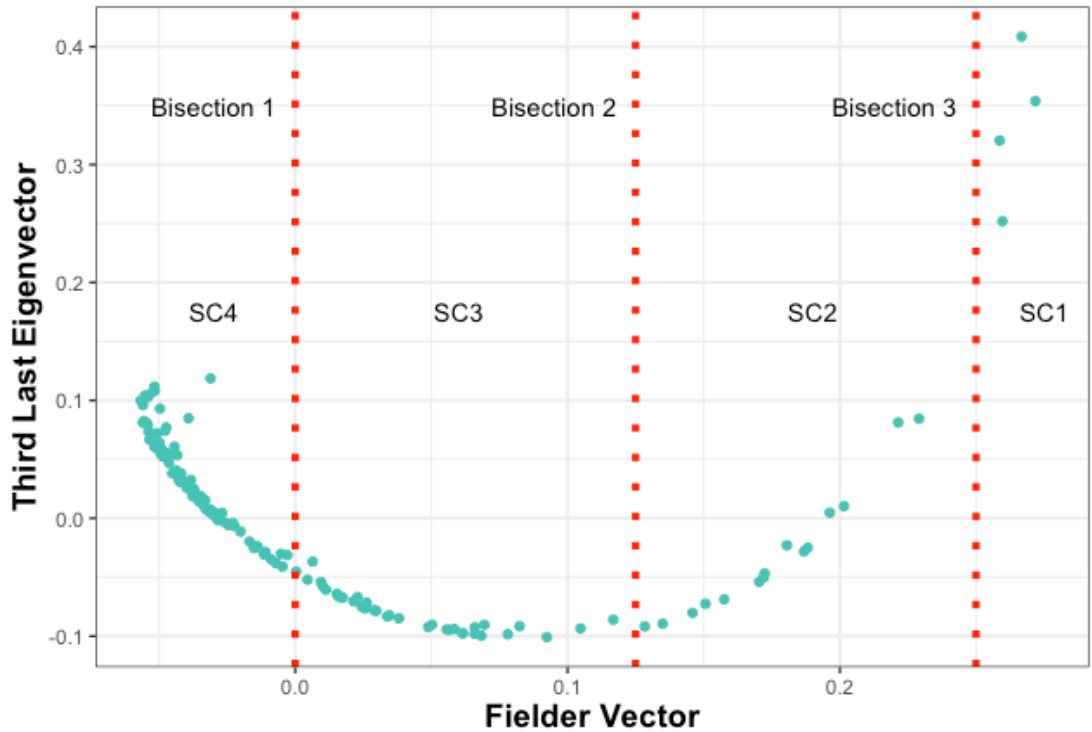

*Note: Red dashed lines show the different bisections using the Feilder vector algorithm*

Figure 8. Silhouette Widths for each cluster

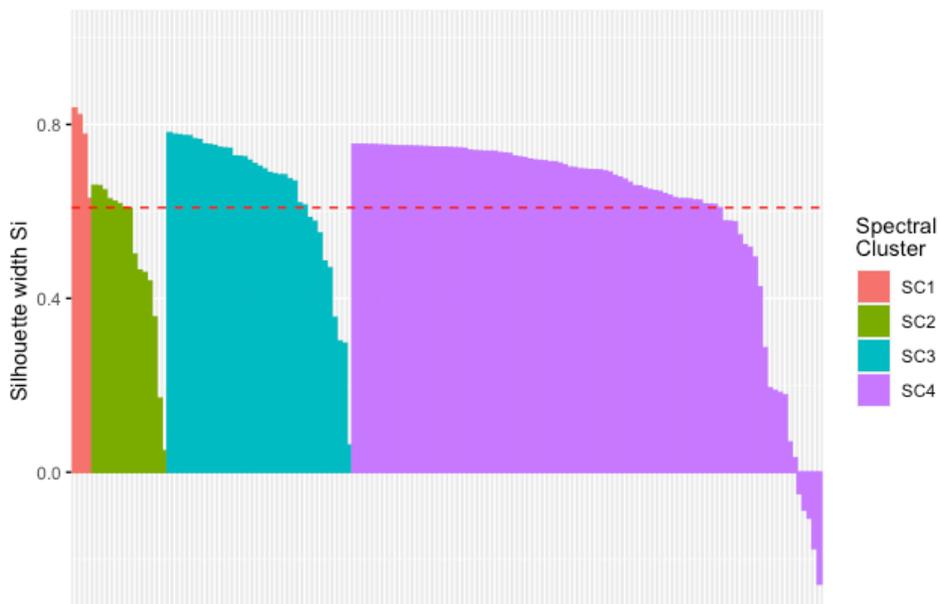

*Note: Red dashed line shows the average Sillouhette Width of 0.61*

Figure 9. Network graph of the Natural Clustering in European Football

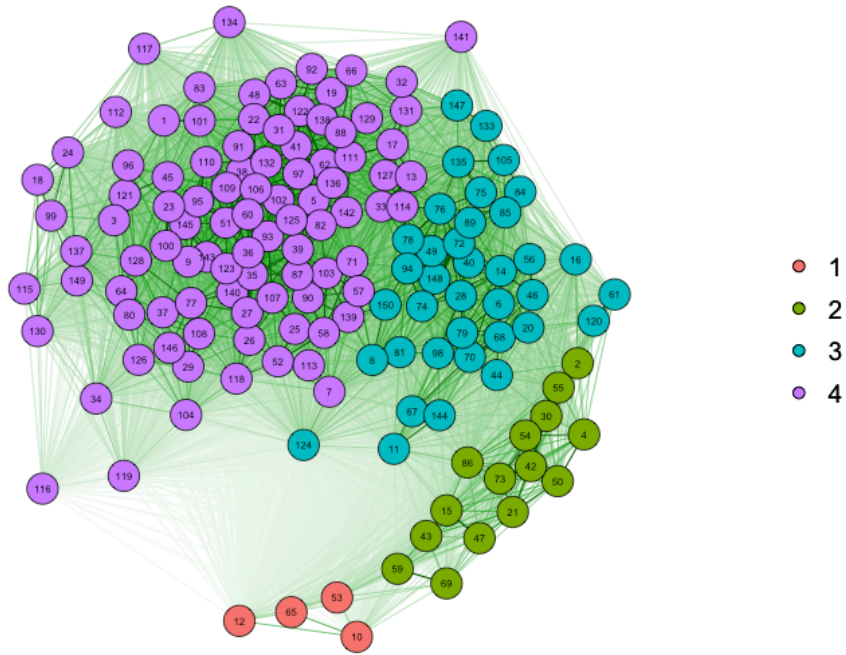

Table 2. The number of teams and

| Cluster | Number of teams | Team ID | Silhouette Index |
| --- | --- | --- | --- |
| SC1 | 4 | 53, 12, 65, 10 | 0.77 |
| SC2 | 15 | 30, 2, 55, 50, 15, 54, 42, 86, 47, 21, 4, 73, 59, 69, 43 | 0.50 |
| SC3 | 37 | 133, 147, 135, 120, 150, 105, 67, 148, 68, 49, 84, 76, 20, 75, 81, 78, 40, 98, 14, 72, 85, 70, 61, 28, 16, 46, 74, 11, 89, 6, 94, 144, 79, 44, 56, 124, 8 | 0.65 |
| SC4 | 94 | 131, 102, 63, 1, 101, 24, 48, 134, 110, 122, 140, 117, 129, 32, 90, 108, 112, 116, 18, 19, 38, 142, 145, 109, 37, 130, 83, 121, 128, 141, 41, 132, 103, 3, 93, 119, 62, 118, 138, 127, 87, 64, 106, 111, 115, 22, 35, 31, 91, 126, 96, 80, 143, 13, 92, 26, 29, 17, 9, 146, 149, 95, 100, 66, 52, 123, 51, 88, 45, 60, 39, 36, 33, 5, 25, 139, 113, 77, 27, 104, 137, 71, 99, 97, 82, 58, 23, 34, 136, 57, 114, 107, 125, 7 | 0.60 |

Using the Fielder vector allows natural groupings of teams (or firms) to be created, and the results show that using the Fielder vector algorithm is relatively effective in finding natural clusters within European football teams. By using unsupervised machine learning and clustering methods, we can objectively identify the dominant teams across Europe. Therefore, clusters 1 and 2 demonstrate the best teams to compete in an elite European Super League – should it be created.

Table 3. The Laplacian eigenvector approach to a new elite European Super League

|    | Team_ID | Team | Tournament | Cluster |
|----|---------|------|------------|---------|
| 1  | 10 | Barcelona | La Liga | SC1 |
| 2  | 12 | Bayern Munich | Bundesliga | SC1 |
| 3  | 53 | Manchester City | Premier League | SC1 |
| 4  | 65 | Paris Saint Germain | Ligue 1 | SC1 |
| 5  | 2  | AC Milan | Serie A | SC2 |
| 6  | 4  | Arsenal | Premier League | SC2 |
| 7  | 15 | Borussia Dortmund | Bundesliga | SC2 |
| 8  | 21 | Chelsea | Premier League | SC2 |
| 9  | 30 | Fiorentina | Serie A | SC2 |
| 10 | 42 | Inter Milan | Serie A | SC2 |
| 11 | 43 | Juventus | Serie A | SC2 |
| 12 | 47 | Liverpool | Premier League | SC2 |
| 13 | 50 | Lyon | Ligue 1 | SC2 |
| 14 | 54 | Manchester United | Premier League | SC2 |
| 15 | 55 | Marseille | Ligue 1 | SC2 |
| 16 | 59 | Napoli | Serie A | SC2 |
| 17 | 69 | Real Madrid | La Liga | SC2 |
| 18 | 73 | Roma | Serie A | SC2 |
| 19 | 86 | Tottenham | Premier League | SC2 |

**Conclusions**

We set out to explore an unsupervised data-driven approach to classifying and identifying dominant teams across Europe's five major football leagues. Our methodology enabled similarities and differences between football teams from disparate leagues to be mapped onto a 2D space. Using a Laplacian eigenmap of the Euclidean distance graph, we have been able to project complex multivariate non-linear relationships. In doing so, we were able to visualise distances between teams, thus identify natural neighbourhoods in which teams inhibit. Through bisection of the Fielder vector, we were able to show how these natural neighbourhoods created suitable clusters to categorise teams. Using the variables that best predict goal difference, we have shown how this approach can identify teams who dominate their respective leagues based on actual performance rather than points. For example, using performance metrics (i.e. shots on target; possession; shots;

shots conceded; and pass success): Barcelona was much closer to Paris Saint-Germain and Bayern Munich than Real Madrid; and Arsenal, Inter Milan and Roma were all closely related.

Indeed, this methodology can be applied to multiple applications within operational research generally. Within a similar context to this paper, the approach can be applied to understanding which players naturally cluster together based on performance metrics to aid decision-making regarding player acquisitions and development. Likewise, it could support merger and acquisition decisions by identifying creditable firms or understanding the impact of strategic choices when creating competitive advantage.

Concerning the specific question of 'who' are the top teams in European soccer, the Laplacian eigenmap methodology classified 15 out of the 16 'breakaway' teams as candidates for the elite league (Der Spiegel, 2018). However, our approach did not select Atlético Madrid and instead selected Napoli, Tottenham Hotspur, Lyon and Fiorentina to the elite European Super League. However, as with any research, there are limitations; firstly, we only consider simple performance metrics, and with the advances in data quality, there are opportunities for more advanced metrics to identify similarities and differences from a performance perspective. Secondly, we do not assess whether the teams selected by bisecting the Fielder vector would make for a competitive league. Thus, further OR research should look to take advantage of more granular and advanced performance data to classify sports teams and forecast what this new elite league will look like with the teams selected here.

Declarations of interest: none.

Appendix 1

| Team_ID | Team | Tournament | Team_ID | Team | Tournament |
|---|---|---|---|---|---|
| 1 | AC Ajaccio | Ligue 1 | 76 | Sassuolo | Serie A |
| 2 | AC Milan | Serie A | 77 | SC Bastia | Ligue 1 |
| 3 | Almeria | La Liga | 78 | Schalke 04 | Bundesliga |
| 4 | Arsenal | Premier League | 79 | Sevilla | La Liga |
| 5 | Aston Villa | Premier League | 80 | Sochaux | Ligue 1 |
| 6 | Atalanta | Serie A | 81 | Southampton | Premier League |
| 7 | Athletic Bilbao | La Liga | 82 | Stoke | Premier League |
| 8 | Atletico Madrid | La Liga | 83 | Sunderland | Premier League |
| 9 | Augsburg | Bundesliga | 84 | Swansea | Premier League |
| 10 | Barcelona | La Liga | 85 | Torino | Serie A |
| 11 | Bayer Leverkusen | Bundesliga | 86 | Tottenham | Premier League |
| 12 | Bayern Munich | Bundesliga | 87 | Toulouse | Ligue 1 |
| 13 | Bologna | Serie A | 88 | Udinese | Serie A |
| 14 | Bordeaux | Ligue 1 | 89 | Valencia | La Liga |
| 15 | Borussia Dortmund | Bundesliga | 90 | Valenciennes | Ligue 1 |
| 16 | Borussia M.Gladbach | Bundesliga | 91 | Valladolid | La Liga |
| 17 | Cagliari | Serie A | 92 | Verona | Serie A |
| 18 | Cardiff | Premier League | 93 | VfB Stuttgart | Bundesliga |
| 19 | Catania | Serie A | 94 | Villarreal | La Liga |
| 20 | Celta Vigo | La Liga | 95 | Werder Bremen | Bundesliga |

| # | Team | League | # | Team | League |
|---|---|---|---|---|---|
| 21 | Chelsea | Premier League | 96 | West Bromwich Albion | Premier League |
| 22 | Chievo | Serie A | 97 | West Ham | Premier League |
| 23 | Crystal Palace | Premier League | 98 | Wolfsburg | Bundesliga |
| 24 | Eintracht Braunschweig | Bundesliga | 99 | Burnley | Premier League |
| 25 | Eintracht Frankfurt | Bundesliga | 100 | Caen | Ligue 1 |
| 26 | Elche | La Liga | 101 | Cesena | Serie A |
| 27 | Espanyol | La Liga | 102 | Cordoba | La Liga |
| 28 | Everton | Premier League | 103 | Deportivo La Coruna | La Liga |
| 29 | Evian Thonon Gaillard | Ligue 1 | 104 | Eibar | La Liga |
| 30 | Fiorentina | Serie A | 105 | Empoli | Serie A |
| 31 | Freiburg | Bundesliga | 106 | FC Koln | Bundesliga |
| 32 | Fulham | Premier League | 107 | Leicester | Premier League |
| 33 | Genoa | Serie A | 108 | Lens | Ligue 1 |
| 34 | Getafe | La Liga | 109 | Metz | Ligue 1 |
| 35 | Granada | La Liga | 110 | Paderborn | Bundesliga |
| 36 | Guingamp | Ligue 1 | 111 | Palermo | Serie A |
| 37 | Hamburger SV | Bundesliga | 112 | Queens Park Rangers | Premier League |
| 38 | Hannover 96 | Bundesliga | 113 | Angers | Ligue 1 |
| 39 | Hertha Berlin | Bundesliga | 114 | Bournemouth | Premier League |
| 40 | Hoffenheim | Bundesliga | 115 | Carpi | Serie A |
| 41 | Hull | Premier League | 116 | Darmstadt | Bundesliga |
| 42 | Inter Milan | Serie A | 117 | Frosinone | Serie A |
| 43 | Juventus | Serie A | 118 | GFC Ajaccio | Ligue 1 |
| 44 | Lazio | Serie A | 119 | Ingolstadt | Bundesliga |
| 45 | Levante | La Liga | 120 | Las Palmas | La Liga |
| 46 | Lille | Ligue 1 | 121 | Sporting Gijon | La Liga |
| 47 | Liverpool | Premier League | 122 | Troyes | Ligue 1 |
| 48 | Livorno | Serie A | 123 | Watford | Premier League |

| # | Team | League | # | Team | League |
|---|---|---|---|---|---|
| 49 | Lorient | Ligue 1 | 124 | RasenBallsport Leipzig | Bundesliga |
| 50 | Lyon | Ligue 1 | 125 | Alaves | La Liga |
| 51 | Mainz 05 | Bundesliga | 126 | Leganes | La Liga |
| 52 | Malaga | La Liga | 127 | Dijon | Ligue 1 |
| 53 | Manchester City | Premier League | 128 | Nancy | Ligue 1 |
| 54 | Manchester United | Premier League | 129 | Middlesbrough | Premier League |
| 55 | Marseille | Ligue 1 | 130 | Crotone | Serie A |
| 56 | Monaco | Ligue 1 | 131 | Pescara | Serie A |
| 57 | Montpellier | Ligue 1 | 132 | Amiens | Ligue 1 |
| 58 | Nantes | Ligue 1 | 133 | Benevento | Serie A |
| 59 | Napoli | Serie A | 134 | Brescia | Serie A |
| 60 | Newcastle United | Premier League | 135 | Brest | Ligue 1 |
| 61 | Nice | Ligue 1 | 136 | Brighton | Premier League |
| 62 | Norwich | Premier League | 137 | Deportivo Alaves | La Liga |
| 63 | Nuernberg | Bundesliga | 138 | Fortuna Duesseldorf | Bundesliga |
| 64 | Osasuna | La Liga | 139 | Girona | La Liga |
| 65 | Paris Saint Germain | Ligue 1 | 140 | Huddersfield | Premier League |
| 66 | Parma | Serie A | 141 | Lecce | Serie A |
| 67 | Rayo Vallecano | La Liga | 142 | Mallorca | La Liga |
| 68 | Real Betis | La Liga | 143 | Nimes | Ligue 1 |
| 69 | Real Madrid | La Liga | 144 | RB Leipzig | Bundesliga |
| 70 | Real Sociedad | La Liga | 145 | SDHuesca | La Liga |
| 71 | Reims | Ligue 1 | 146 | Sheffield United | Premier League |
| 72 | Rennes | Ligue 1 | 147 | SPAL | Serie A |
| 73 | Roma | Serie A | 148 | Strasbourg | Ligue 1 |
| 74 | Saint-Etienne | Ligue 1 | 149 | Union Berlin | Bundesliga |
| 75 | Sampdoria | Serie A | 150 | Wolverhampton Wanderers | Premier League |